\documentclass{aa}  

\usepackage{graphicx}
\usepackage{txfonts}
\usepackage{epsfig}
\usepackage{longtable}
\begin{document}

   \title{Brightness variations of the FUor-type eruptive star V346
     Nor\footnote{Based on observations collected at the European
       Organisation for Astronomical Research in the Southern
       Hemisphere under ESO programmes 71.C-0526(A), 179.B-2002,
       and 381.C-0241(A).}}

   \author{\'A. K\'osp\'al\inst{1}
          \and
          P. \'Abrah\'am\inst{1}
          \and
          Ch. Westhues\inst{2}
          \and
          M. Haas\inst{2}
          }

   \institute{Konkoly Observatory, Research Centre for Astronomy and
     Earth Sciences, Hungarian Academy of Sciences, 
     PO Box 67, 1525 Budapest, Hungary
     \email{kospal@konkoly.hu}
     \and
     Astronomisches Institut, Ruhr-Universit\"at Bochum,
     Universit\"atsstra\ss{}e 150, D-44801 Bochum, Germany
     }

   \date{Received date; accepted date}

%  \abstract

\abstract{Decades after the beginning of its FU Orionis-type outburst,
  V346~Nor unexpectedly underwent a fading event of $\Delta{}K$ =
  4.6\,mag around 2010. We obtained near-infrared observations and
  re-analysed data from the VISTA/VVV survey to outline the brightness
  evolution. In our VLT/NaCO images, we discovered a halo of scattered
  light around V346~Nor with a size of about 0$\farcs$04 (30 au). The
  VISTA data outlined a well-defined minimum in the light curve at
  late 2010/early 2011, and tentatively revealed a small-amplitude
  periodic modulation of 58\,days. Our latest data points from 2016
  demonstrate that the source is still brightening but has not reached
  the 2008 level yet. We used a simple accretion disk model with
  varying accretion rate and line-of-sight extinction to reproduce the
  observed near-infrared magnitudes and colors. We found that before
  2008, the flux changes of V346~Nor were caused by a correlated
  change of extinction and accretion rate, while the minimum around
  2010 was mostly due to decreasing accretion. The source reached a
  maximal accretion rate of ${\approx}10^{-4}\,M_{\odot}$\,yr$^{-1}$
  in 1992. A combination of accretion and extinction changes was
  already invoked in the literature to interpret the flux variations
  of certain embedded young eruptive stars.}


\titlerunning{Brightness variations of V346 Nor}
\authorrunning{K\'osp\'al et al.}

   \maketitle

%-----------------------------------------------------------------
% INTRODUCTION
%-----------------------------------------------------------------
\section{Introduction}

\begin{figure}
\centering \includegraphics[width=7.5cm,angle=0]{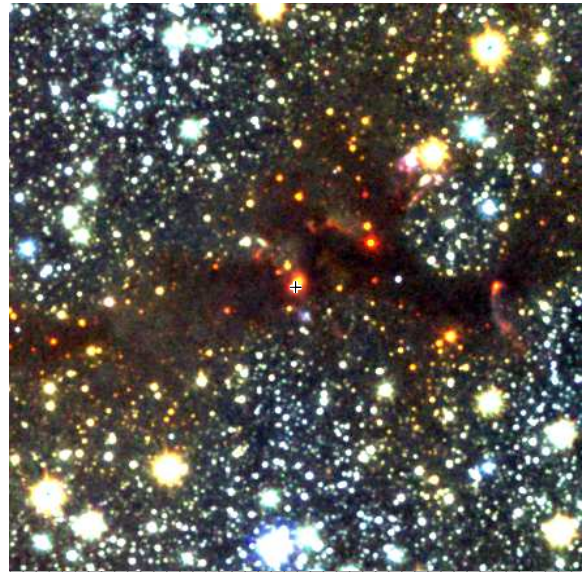}
\caption{V346~Nor (black {\it plus} sign in the center) and its
  surroundings in a $JHK_{\rm S}$ color composit image. The
  observations were taken within the VISTA Variables in The Via Lactea
  (VVV) Survey on March 15, 2010. North is up and east is to the
  left. The displayed area is 2$\farcm$5${\times}$2$\farcm$5 in
  size. The dark lane across the image is due to extinction by the
  Sa~187 molecular cloud.}
\label{fig:composit}
\end{figure}

FU Orionis-type stars (FUors) are low-mass pre-main sequence objects
characterized by 4-6\,magnitude optical outbursts due to temporarily
enhanced accretion from the circumstellar disk to the star
\citep{hk96}. Following the outburst of the first such object, FU~Ori
in 1937, now more than two dozen FUors and FUor candidates are known
\citep{audard2014}. V346~Nor was discovered by \citet{elias1980} as a
source within a few arcseconds of the HH~57 nebulosity, the latter
being a faint, compact H$\alpha$ emitting knot \citep{schwartz1977},
located in the Sa~187 molecular cloud within the Norma 1 association,
at a distance of 700\,pc (\citealt{reipurth1981}, see also
Fig.~\ref{fig:composit}). A few years later \citet{graham1983}
reported the appearance of a star-like source at the northeastern tip
of HH~57, probably coinciding with the source in
\citet{elias1980}. They mentioned that the star was not visible in
1976 \citep{schwartz1977}, but a diffuse patch is clearly discernible
in the blue plates of the ESO/SERC Sky Survey, obtained in April--June
1975 \citep{holmberg1974,reipurth1981}. Therefore, V346~Nor
transformed from a faint diffuse nebula to a bright point-source some
time between 1976 and 1980. Based on this and on the spectroscopic
properties of the star, \citet{reipurth1983a} suggested that V346~Nor
was undergoing a FUor-type outburst. \citet{frogel1983} presented
photometry from 2.2 to 20$\,\mu$m and remarked that the colors of the
object are similar to those of FU~Ori and V1057~Cyg.

Subsequent near-infrared (near-IR) photometry indicated that V346~Nor
was gradually brightening in the $K$ band, reaching a broad maximum
between about 1990 and 2000 (\citealt{frogel1983, reipurth1983b,
  reipurth1985, kh1991, molinari1993, prusti1993, reipurth1997,
  abraham2004}, see also Fig.~\ref{fig:light}). \citet{kh1991}
presented the broad-band optical-IR spectral energy distribution (SED)
of V346~Nor, while \citet{weintraub1991} published submillimeter and
millimeter photometry for it. Both groups concluded that the object is
surrounded by a significant amount of circumstellar material, in the
form of an actively accreting disk and a flattened envelope. Recently,
\citet{kraus2016} reported a dramatic brightness decrease of V346~Nor
and a subsequent brightening. They interpret these results as a 2-3
orders of magnitude drop in the accretion rate, followed by the onset
of a new outburst. In order to better understand this spectacular
event, and to follow up the evolution of the system, we present new
near-IR observations of V346~Nor, and re-evaluate the data taken by
the VISTA telescope. We analyze the brightness and color variations
observed in V346~Nor and compare the results with similar fading
events of highly accreting young stellar objects from the literature.

%-----------------------------------------------------------------
% OBSERVATIONS
%-----------------------------------------------------------------
\section{Observations, data reduction, and photometry}

\begin{figure}
\centering \includegraphics[width=8.5cm,angle=0]{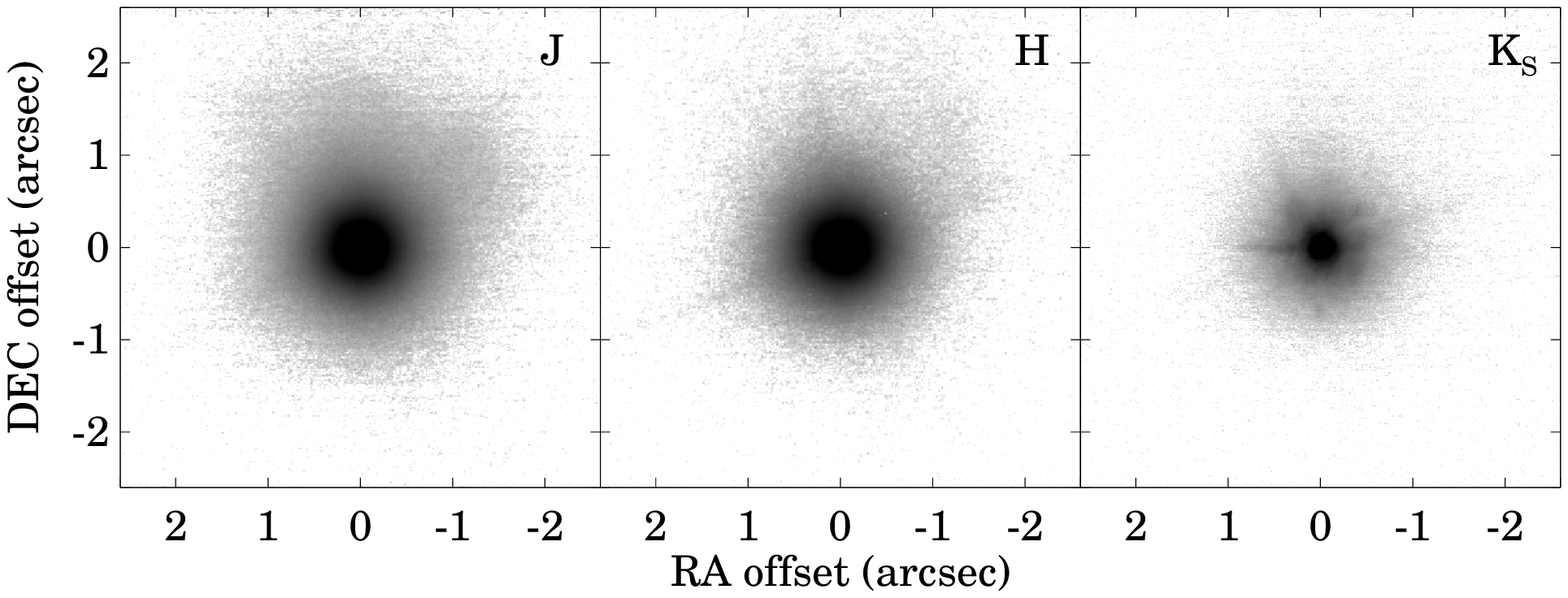}
\centering \includegraphics[width=8.5cm,angle=0]{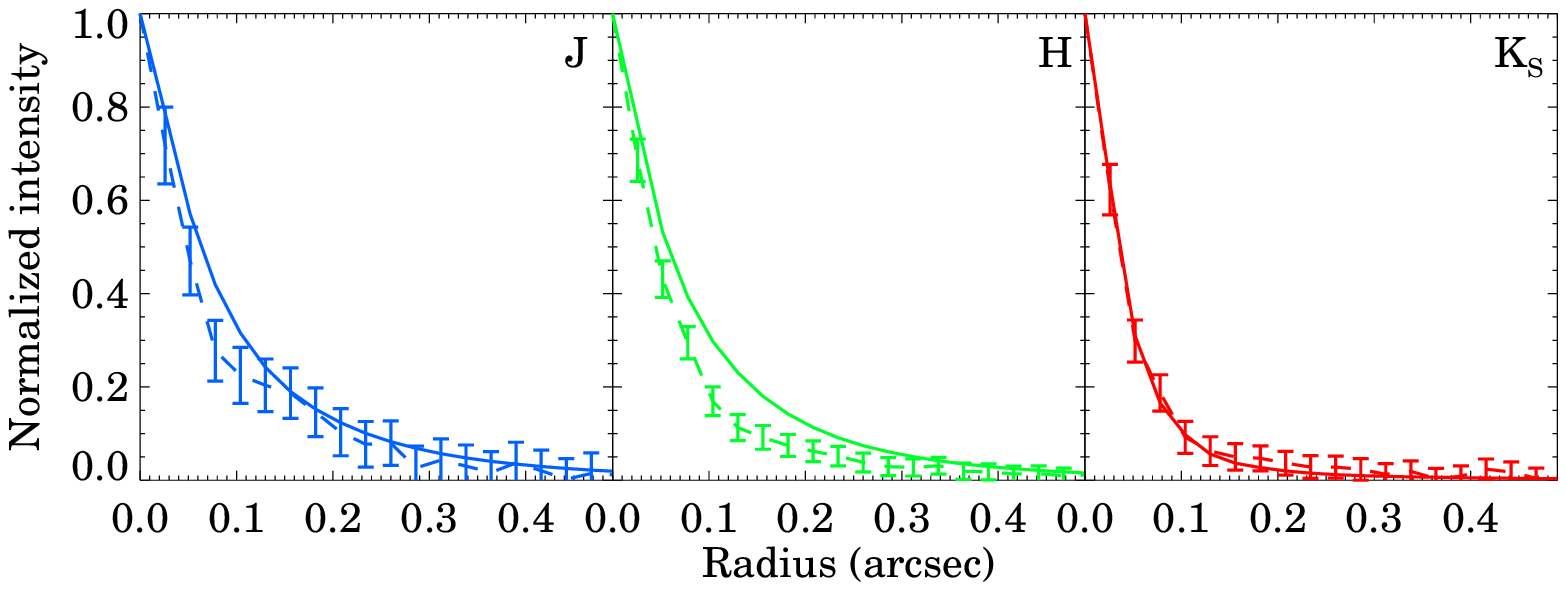}
\caption{{\it Top:} VLT/NaCo $J$, $H$, and $K_{\rm S}$ images of
  V346~Nor from 2008. The color scale is logarithmic. {\it Bottom:}
  Radial intensity profiles of V346~Nor (solid curves) and of another,
  fainter star visible in the field of view (dashed curves) measured
  in our NaCo $J$, $H$, and $K_{\rm S}$ images.  The uncertainty of
  the brightness profile of the fainter star is indicated by error
  bars, while the uncertainty of V346~Nor's profile is less than the
  curve thickness.}
\label{fig:naco}
\end{figure}

We observed V346~Nor with the NaCo adaptive optics instrument on the
UT4 of European Southern Observatory's Very Large Telescope (VLT) at
Cerro Paranal, Chile, on April 10/11, 2008, as part of project
381.C-0241 (PI: \'A.~K\'osp\'al). The weather conditions were good and
the typical optical seeing was around 1$''$. We obtained $J$, $H$, and
$K_{\rm S}$-band images with the N20C80 dichroic and the
13\,mas\,pixel$^{-1}$ scale camera. We observed
2MASS~J16323308$-$4457314 as a photometric standard with the same
instrumental setup as the science target. This star is only 2$'$ away
from V346~Nor and has similar 2MASS magnitudes ($J$=10.288\,mag,
$H$=8.395\,mag, $K_{\rm S}$=7.242\,mag). Our NaCo images are displayed
in Fig.~\ref{fig:naco}. We obtained aperture photometry for both
V346~Nor and the photometric standard using an aperture radius of
2$''$ and sky annulus between 2$\farcs$6 and 3$\farcs$9. This large
aperture was chosen to include all the flux of V346~Nor, which appears
slightly extended in our NaCo images (see below), therefore, our
photometry can be compared to earlier unresolved photometry from the
literature. The obtained instrumental magnitudes were converted to
standard magnitudes using the 2MASS values for the photometric
standard. We also downloaded archival $J$ and $K_{\rm S}$-band
  NaCo observations from June 12, 2003, and reduced and extracted
  photometry from them in a similar way. The obtained brightnesses of
V346~Nor are presented in Tab.~\ref{tab:phot}.

Our group observed the area around V346~Nor with the InfraRed Imaging
System (IRIS) at Bochum Observatory near Cerro Armazones. IRIS is a
80\,cm telescope equipped with a 1k$\times$1k infrared camera. The
system provides a resolution of 0.74$''$/pixel and a field-of-view of
13$'{\times}$13$'$. Data were taken between June 26, 2010 and July 1,
2010, as well as between August 22 and 25, 2016, in the $J$, $H$ and
$K_{\rm S}$ bands. Individual frames were combined to eliminate the
sky signal and correct for flatfield differences. V346~Nor was not
visible in the $J$ band, but was detected in all $H$ and $K_{\rm S}$
frames. All $J$-band images obtained on the same night were combined
into mosaics, and 3\,$\sigma$ upper limits were determined. For the
other filters, we performed aperture photometry using the same
aperture and sky annulus sizes as for the NaCo images. For the
photometric calibration, we used a set of about 50 2MASS stars with
quality flag `A' to determine the offset between the instrumental and
the 2MASS magnitudes. We found that no color term was needed for this
transformation. The uncertainty of the final photometry is the
quadratic sum of the formal uncertainty of the aperture photometry and
the photometric calibration. The resulting $J$ upper limits and
$HK_{\rm S}$ magnitudes are presented in Tab.~\ref{tab:phot}.

We observed V346~Nor with the SMARTS 1.3\,m telescope at Cerro Tololo
on June 7 and August 9, 2016. The telescope is equipped with the
ANDICAM instrument, which provides simultaneous optical and IR
images. The CCD for the ANDICAM is a Fairchild 447 2k$\times$2k chip,
which we used with 2$\times$2 binning, resulting in a binned pixel
scale of 0.371$''$/pixel, and a field of view of about
6$'\times$6$'$. The IR Array for the ANDICAM is a Rockwell
1k$\times$1k HgCdTe ``Hawaii'' Array, also used with 2$\times$2
binning, with 0.276$''$/pixel binned scale and
2$\farcm$4$\times$2$\farcm$4 field of view. We used the
Johnson-Kron-Cousins $VRI$ optical and CIT/CTIO $JHK$ IR filters. A
5-point dithering was done to enable bad pixel removal and sky
subtraction in the IR images. Bias and flat correction for the optical
images were done by the Yale SMARTS team. Although HH~57 is faintly
visible in our $V$ and $R$ images, V346~Nor itself is not detected in
the optical. We used magnitude values from the UCAC4 catalog
\citep{zacharias2013} to calibrate the images and determined 3$\sigma$
upper limits of $V\,{>}\,$20.8\,mag, $R\,{>}\,$19.6\,mag, and
$I\,{>}\,$18.8\,mag for V346~Nor. In the near-IR regime, V346~Nor was
detected in all three bands, and we performed photometry the same way
as described for the IRIS images. The resulting $JHK$ magnitudes for
V346~Nor are presented in Tab.~\ref{tab:phot}.

V346~Nor was covered as part of the VISTA Variables in the Via Lactea
Survey, an ESO public survey using the VISTA 4.1\,m telescope and the
VIRCAM near-IR camera \citep{minniti2010}. We downloaded all VIRCAM
images from this survey covering V346~Nor. To obtain photometry that
can be compared with our NaCO, IRIS, and SMARTS data, we performed our
own flux extraction with the same aperture as described above. At some
epochs, our values differ from those published by \citet{kraus2016}
for the same measurements. The reason is probably a different
treatment of the known non-linearity of the VIRCAM detectors for
bright sources \citep[e.g.,][]{saito2012}. The correction we applied
is described in details in Appendix~\ref{sec:appendix_a}. The $J$ and
$H$ photometry, as well as the $Ks$-band results after the
non-linearity correction are also given in Tab.~\ref{tab:phot}.

%-----------------------------------------------------------------
% RESULTS
%-----------------------------------------------------------------
\section{Results}

The top part of Figure~\ref{fig:naco} shows our NaCo $JHK_{\rm S}$
images from 2008 of V346~Nor, while the bottom panel displays the
normalized, azimuthally averaged radial brightness distributions of
V346~Nor, and another, fainter star visible in the field of view at a
distance of 5$\farcs$7, position angle of 13$^{\circ}$ east of
north. Assuming that this nearby faint star is a point source, the
comparison of the brightness profiles show that V346~Nor is extended
in the $J$ and $H$ bands, while it is consistent with a point source
in the $K_{\rm S}$ band. Deconvolved sizes using Gaussian
deconvolution are 0$\farcs$041$\pm$0$\farcs$016 (29$\pm$11\,au) in
$J$, 0$\farcs$037$\pm$0$\farcs$010 (25$\pm$7\,au) in $H$, and we can
give an upper limit of 0$\farcs$02 (13\,au) for the $K_{\rm S}$-band
size. The size of the near-IR emitting region in circumstellar disks
is typically only a few au, therefore, the emission seen in the NaCo
images has to be scattered light. The $J$ and $H$ images are slightly
asymmetric with the northern part slightly more extended than the
southern part. No such asymmetry is evident in the $K_{\rm S}$ image.

\begin{figure}
\centering
\includegraphics[width=8.5cm,angle=0]{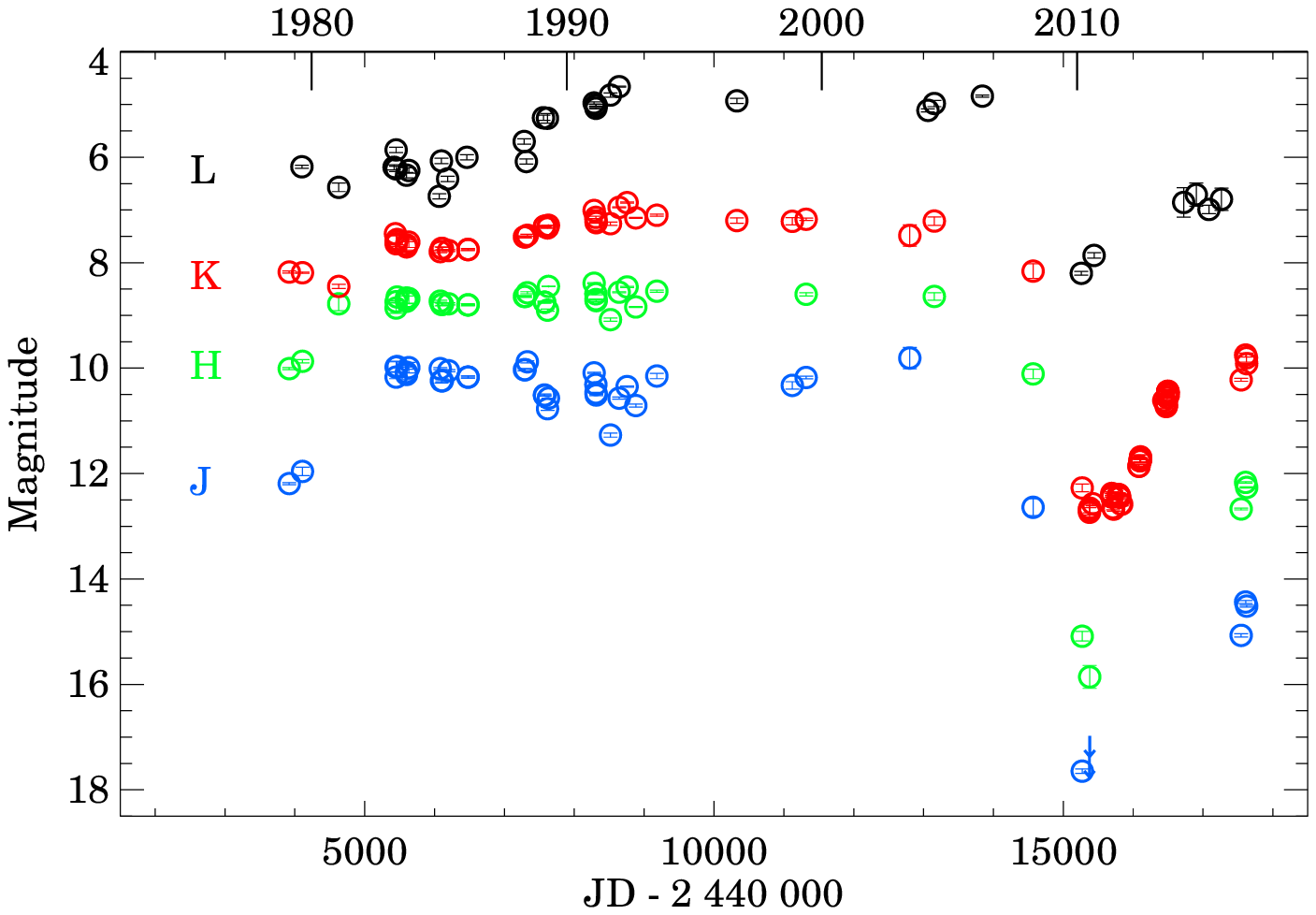}
\includegraphics[width=8.5cm,angle=0]{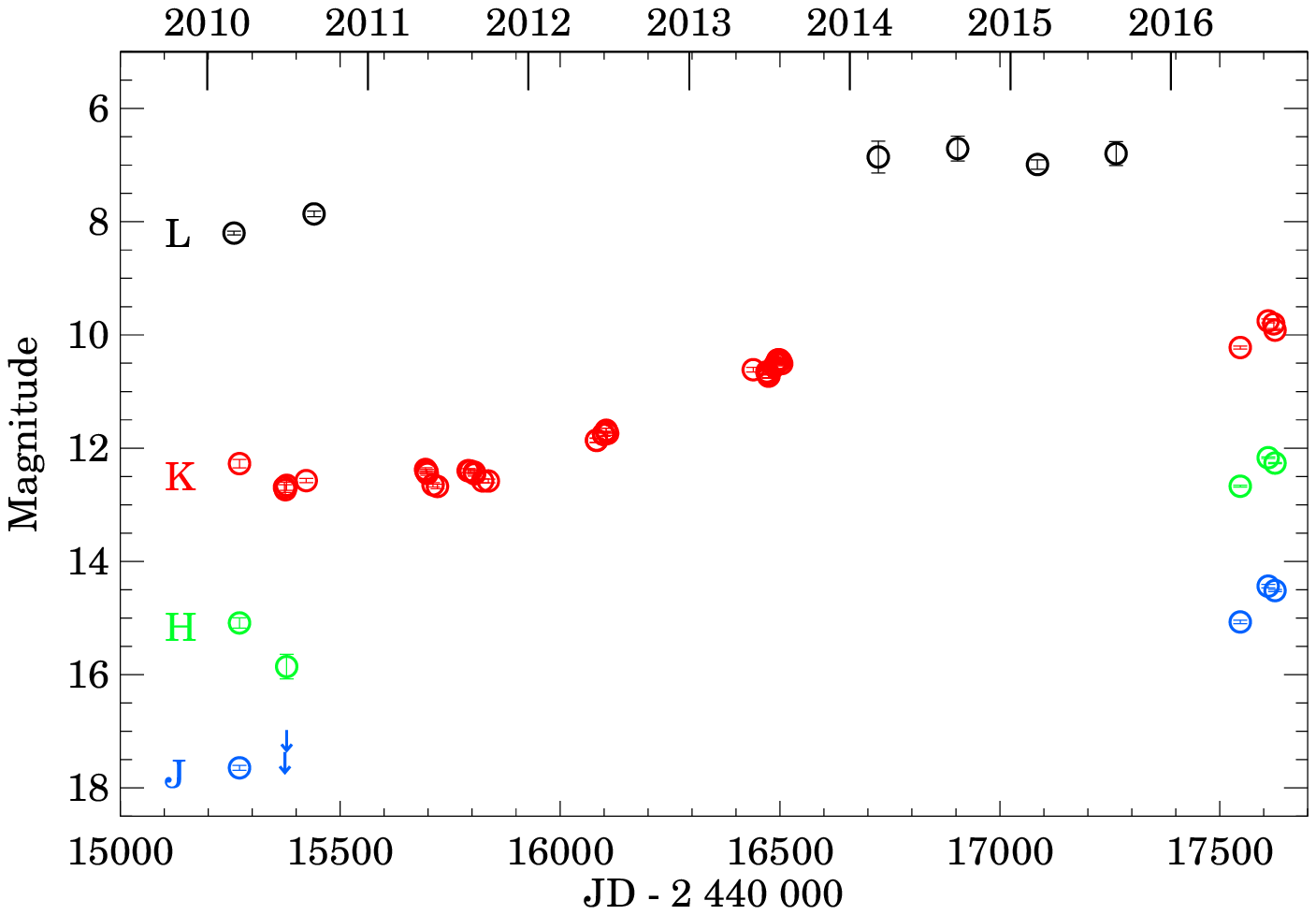}
\caption{Near-IR light curves of V346~Nor. Data points are from
  \citet{elias1980, frogel1983, graham1985, reipurth1983a,
    reipurth1983b, reipurth1985, kh1991, molinari1993, prusti1993,
    reipurth1997, quanz2007, connelley2008}, the 2MASS, DENIS, AllWISE
  and NEOWISE catalogs \citep{cutri2003, cutri2013, cutri2015}, and
  this work. Downward arrows indicate upper limits.}
\label{fig:light}
\end{figure}

Figure~\ref{fig:light} shows the near-IR light curves of V346~Nor. The
first few data points indicate a brightening between 1979 and 1983,
followed by a relatively constant period until about 1988. Afterwards,
the $K$ and $L$-band data show a gradual brightening until 1992,
already reported in \citet{abraham2004}. The $J$ and $H$ light curves
were rather flat, with small, $<$\,1\,mag brightness variations.
After 2003-2004 the source significantly faded, especially at short
wavelengths and in 2008 displayed similar brightness although slightly
redder color than the first $JHK$ photometry in 1979.

Some time around 2008, V346~Nor started a dramatic fading, and reached
a minimum around late 2010-early 2011 (Fig.~\ref{fig:light},
bottom). Afterwards, the source quickly brightened by
$\Delta{}K\,=\,$2.4\,mag in about 3 years, and another
$\Delta{}K\,=\,$0.6\,mag in the following 3 years, indicating a slower
brightening rate. As of 2016 August, V346~Nor has not yet reached the
brightness level it displayed in 1980--2000. This is well visible in
the near-IR light curves, but our optical upper limits also supports
this, as the star would have been visible in our $VRI$ images had it
been as bright as between 1980--2000. The deep minimum was also
visible in the WISE 3.4$\,\mu$m photometry, although with smaller
amplitude.

The lower panel of Fig.~\ref{fig:light} shows that the minimum in the
$K$-band has a parabola-like light curve shape. By fitting and
subtracting a second-order polinomial from the photometry between 2010
and 2014, the obtained residuals are on the order of 0.2\,mag, and
suggest a possible periodic modulation. We calculated a Lomb-Scargle
periodogram for the residuals, and found a tentative 58$\pm$2\,day
period in the data with a false alarm probability of
3$\times$10$^{-3}$ (see also Fig.~\ref{fig:periodogram} in
Appendix~\ref{sec:appendix_b}). Similar periodicities in the light
curves were already found in, e.g., V1647~Ori, where it was explained
by an orbiting dust cloud \citep{acosta2007}, and in V960\,Mon, where
is was explained by a putative close companion \citep{hackstein2015}.

\begin{figure}
\centering
\includegraphics[width=8.5cm,angle=0]{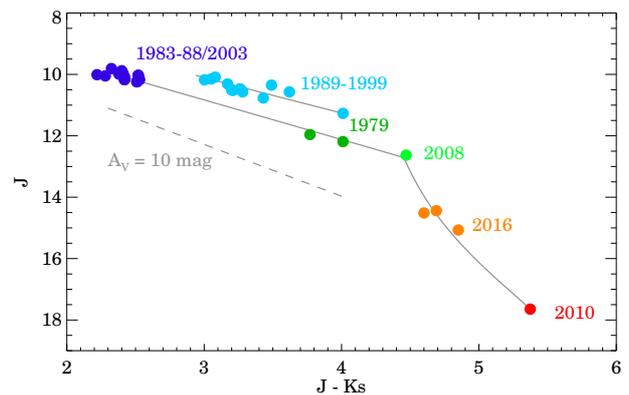}
\caption{Near-IR color-magnitude diagram for V346~Nor. The reddening
  path is marked with dashed line \citep{cardelli1989}. The solid gray
  curves indicate our reddened accretion disk model fits (see details
  in text).}
\label{fig:tcd}
\end{figure}

%-----------------------------------------------------------------
% DISCUSSION
%-----------------------------------------------------------------
\section{Discussion and Conclusions}

The color-magnitude diagram in Fig.~\ref{fig:tcd} shows that all data
points until 2008 form an approximately linear strip, while the later
measurements deviate from this trend, suggesting different physical
mechanisms for the brightness and color changes before and after
2008. In order to reproduce the observations, in each epoch we fitted
the $JHK_{\rm S}$ data points by a steady, optically thick,
geometrically thin, viscous accretion disk, with radially constant
mass accretion rate. Such disk models were successfully proposed and
used to reproduce the SEDs of FUors by \citet{hk96}, \citet{zhu2007},
and \citet{kospal2016}. We calculated the disk's SED by integrating
the fluxes of concentric annuli between the stellar radius and R$_{\rm
  out}$, assuming blackbody emission. We reddened the model fluxes by
different $A_{V}$ values, using the extinction law from
\citet{savage1979} with $R_V$=3.1. Fixing the outer radius of the
accretion disk (R$_{\rm out}$ = 2\,au, the exact value has a
negligible effect on the near-IR fluxes), we have only two free
parameters, the product of the stellar mass and the accretion rate
$M\dot{M}$, and the extinction $A_{V}$. We fixed the stellar mass and
radius to typical low-mass YSO values of 1$\,M_{\odot}$ and
3.0$\,R_{\odot}$ (the resulting accretion rate is inversely
proportional to the adopted stellar mass). The inclination of the
V346~Nor disk is not known, thus we adopted
2/$\pi\,{\approx}\,$40$^{\circ}$, the mean expected value if the disk
is randomly oriented. The fitting procedure was performed with
$\chi^2$ minimization. More extreme extinction laws (up to $R_V$=5.3,
\citealt{cardelli1989}) would change the fitted $A_{V}$ and $\dot{M}$
values by less than 15\%, which is within the formal uncertainty of
our fitting procedure.

The bluest points in Fig.~\ref{fig:tcd} correspond to measurements
obtained in 1983-88 and 2003, when the system exhibited approximately
the same brightness and color. We found that these points can be well
fitted by our disk model with $\dot{M}$ =
1.0$\times10^{-5}\,M_{\odot}$\,yr$^{-1}$ and line-of-sight reddening
of $A_V$=6.7\,mag. We obtained similarly good fits for the 1979 and
2008 SEDs with $\dot{M}$ = 2.1$\times10^{-5}\,M_{\odot}$\,yr$^{-1}$
and $\dot{M}$ = 4.5$\times10^{-5}\,M_{\odot}$\,yr$^{-1}$,
respectively. The reddening, however, was significantly higher,
16.7\,mag in 1979 and 21.5\,mag in 2008. We simulated the time
evolution of the system by computing a sequence of models of gradually
changing $A_V$ from 21.5 to 6.7~mag and $\dot{M}$ from
4.5$\times10^{-5}$ to 1.0$\times10^{-5}\,M_{\odot}$\,yr$^{-1}$. The
resulting line is plotted in Fig.~\ref{fig:tcd}. For most epochs, our
disk model fits the SED shape well with typical formal uncertainties
of 1-2\,mag in $A_V$ and 10-30\% in $\dot{M}$. After the minimum in
2010, the shape of the SEDs can be reproduced less well with the
accretion disk model, resulting in formal uncertainties of 6-8\,mag in
$A_V$, and up to a factor of 6 in $\dot{M}$.

The data points obtained between 1989 and 1999 are situated above the
model line. These SEDs can also be fitted with our disk model, but
with higher $\dot{M}$ values than at any time before. $A_V$ was
between 12.1 and 19.2~mag in this period. In particular, we found that
the accretion rate showed a maximum in 1992 January, with
$\dot{M}$=9.8$\times10^{-5}$ and $A_V$=16.8~mag. Here, we again
computed a sequence of models with gradually changing $A_V$ from 19.2
to 12.1~mag and $\dot{M}$ from 9.8$\times10^{-5}$ to
3.5$\times10^{-5}\,M_{\odot}$\,yr$^{-1}$, also plotted in
Fig.~\ref{fig:tcd}.
 
\citet{kraus2016} suggested that the minimum around 2010 is related to
a large drop in the accretion rate. Using our accretion disk model, we
found that by keeping a constant $A_V$=21.5\,mag between 2008 and
2010, and adding an accretion disk model of gradually increasing
$\dot{M}$ to the SED measured in 2010 we can reach the 2008 data point
(Fig.~\ref{fig:tcd}). In the minimum, the measured fluxes constrain
the model accretion rate below 4$\times10^{-7}\,M_{\odot}$\,yr$^{-1}$,
thus the change in accretion rate was at least a factor of 100 or
more, in agreement with the findings of \citet{kraus2016}. In our
modeling, the scattered light component, indicated by our NaCo
observations in 2008, was not included, since its consistent treatment
would be beyond the scope of this Letter.

Our results demonstrated that while the rapid fading in 2010 was an
accretion event, the flux evolution beforehand was due to a correlated
change in extinction and accretion rate together. That is, increasing
accretion rate is accompanied by growing extinction towards the
source. V346~Nor is similar to a group of highly variable young
stellar objects whose flux changes are due to a combined effect of
changing accretion rate and variable circumstellar extinction. Such
objects are, e.g., H$\alpha$11, PV~Cep, V1647~Ori, and V899~Mon
\citep{kun2011b,kun2011,mosoni2013,ninan2015}. V346~Nor also resembles
the young eruptive star V2492~Cyg in several aspects: the pre-outburst
position of V2492~Cyg in the near-IR color-color diagram is close to
the point of the 2010 minimum of V346~Nor, it also underwent a large
accretion change at the beginning of its outburst, and in the high
state, the line-of-sight extinction is continuously varying
\citep{kospal2011,kospal2013,hillenbrand2013}. Therefore, the observed
variability in both V346~Nor and V2492~Cyg are governed by a
combination of changing accretion and extinction.

The minimum of V346~Nor in 2010 was immediately followed by the
re-brightening of the source. As of 2016, the source is on its way
back to its 2008 state in the color-magnitude diagram
(Fig.~\ref{fig:tcd}). This suggests that the brightening is governed
by the same process that was responsible for the fading, namely
changing accretion at a constant high extinction. Indeed, our simple
accretion disk model can reproduce the $JHK_{\rm S}$ fluxes measured
in 2016 by assuming $\dot{M} =$
3.3$\times10^{-6}\,M_{\odot}$\,yr$^{-1}$, and $A_V$ = 19.8\,mag. The
brightening of the source is still ongoing, and our latest
observations suggest that after a relatively constant period after
2013, V346~Nor might have entered again a fast brightening phase.

In the classical FUor outburst models, the eruption ends when the
inner disk completely depletes, and a new eruption can start only when
the disk material is replenished, typically in several thousand or ten
thousand years \citep{bell1994}. Therefore, the relatively short
minimum of V346~Nor is unlikely to signal the end of the large FUor
outburst in the classical sense. It was more likely a temporary halt
of the accretion onto the star. A similar phenomenon was observed in
V899~Mon by \citet{ninan2015}, and in V1647~Ori, where the 2004--2006
outburst finished and then restarted a few years later. The physical
mechanism of this temporary stop is not known yet. \citet{ninan2015}
speculated about several possible explanations, with the constraint
that these processes should be able to return the system to the
pre-fading state within a few years. The detailed understanding of the
evolution of the V346~Nor outburst will bring us closer to the
understanding of these enigmatic sudden fadings of eruptive stars.

%-----------------------------------------------------------------
% ACKNOWLEDGEMENTS
%-----------------------------------------------------------------
\begin{acknowledgements}The authors thank the referee for his/her
  useful comments. This work was supported by the Momentum grant of
  the MTA CSFK Lend\"ulet Disk Research Group, and by the NKFIH
  research fund OTKA\,101393.
\end{acknowledgements}

%-----------------------------------------------------------------
% BIBLIOGRAPHY
%-----------------------------------------------------------------
\bibliographystyle{aa}
%\bibliography{paper}{}

%-----------------------------------------------------------------
% ONLINE MATERIAL
%-----------------------------------------------------------------

\begin{appendix}

\section{Near-IR photometry of V346~Nor}
\label{sec:appendix_a}

\begin{figure}
\centering
\includegraphics[width=8.5cm,angle=0]{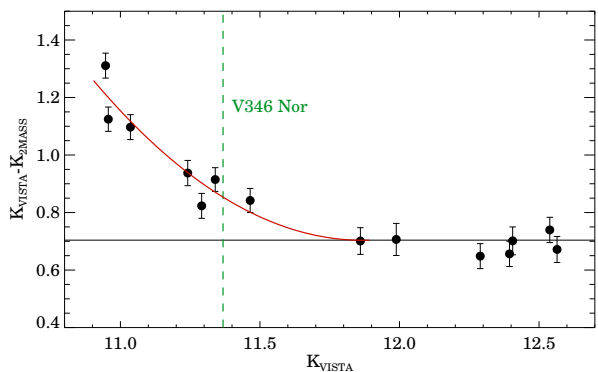}
\caption{Demonstration of the non-linearity correction applied for the
  photometry of V346~Nor.}
\label{fig:nonlinearity1}
\end{figure}

\begin{figure}
\centering
\includegraphics[width=8.5cm,angle=0]{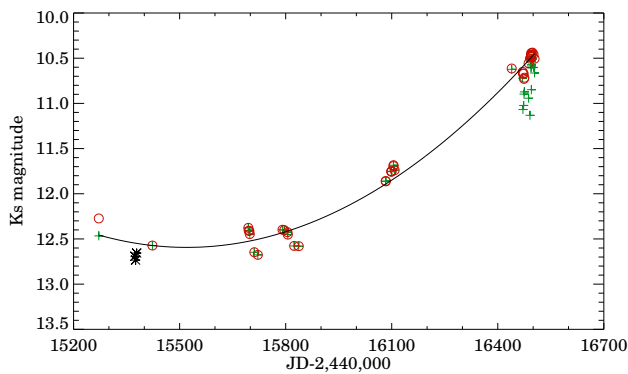}
\caption{VISTA/VIRCAM photometry of V346~Nor without (green plus
  signs) and with (red circles) correction for
  non-linearity. Asterisks indicate our IRIS photometry. The black
  solid curve is a parabola fitted to the data points to remove the
  long-term trend before the period analysis (see
  Sec.~\ref{sec:appendix_b}).}
\label{fig:nonlinearity2}
\end{figure}

The comparison of our aperture photometry with the 2MASS catalog
revealed the known detector issue that stars brighter than about
11\,mag in the $K_{\rm S}$ band enter the non-linear regime of the
VIRCAM detector \citep{saito2012}. While close to its minimum V346~Nor
was below this limit, after about May 2013 it became brighter than
11\,mag. In order to correct for the underestimation of the signal, we
plotted the offsets between the instrumental and the 2MASS magnitudes
for all stars with quality flag `A' in the image as a function of the
instrumental magnitude. We fitted the distribution of points with a
first or second order polynomial, and determined the offset valid for
V346~Nor from this fit (for an example, see
Fig.~\ref{fig:nonlinearity1}). The necessary correction due to the
non-linearity was typically in the 0.1--0.2 mag range, with a few
higher values up to 0.4--0.6\,mag. Fig.~\ref{fig:nonlinearity2} shows
the $K_{\rm S}$-band light curve without and with the non-linearity
correction, demonstrating that our correction significantly reduced
the scatter of the data points obtained close in time.

\section{Period analysis}
\label{sec:appendix_b}

\begin{figure}
\centering
\includegraphics[width=8.5cm,angle=0]{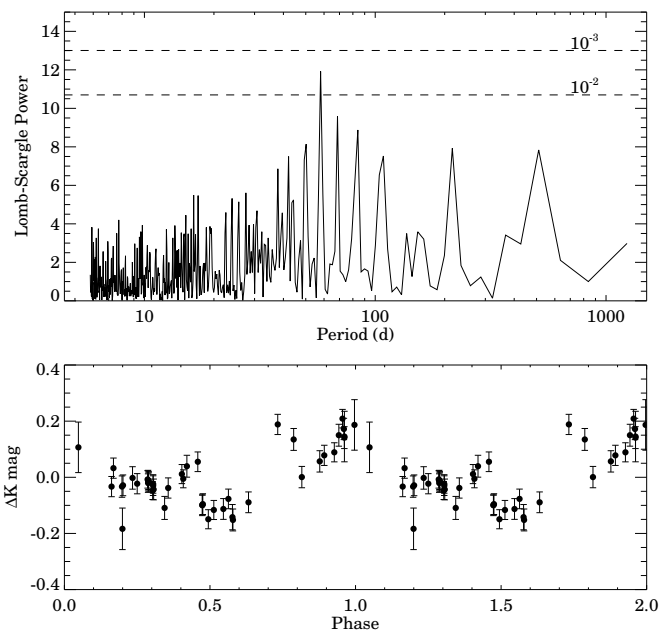}
\caption{{\it Top:} Lomb-Scargle periodogram of the light curve of
  V346~Nor after removing a parabolic trend, as illustrated in
  Fig.~\ref{fig:nonlinearity2}. The highest peak corresponds to a
  period of 58~days. The dashed lines show the powers corresponding to
  false alarm probabilities of 10$^{-3}$ and 10$^{-2}$. {\it Bottom:}
  Phase-folded light curve showing the data points after removing the
  parabolic trend and folded with a period of 58~days.}
\label{fig:periodogram}
\end{figure}

\begin{table*}
\caption{Near-IR photometry of V346~Nor.}\label{tab:phot}
\begin{tabular}{ccccccc}
\hline \hline
Date       & JD$\,{-}\,$2\,400\,000 & $J$            & $H$              & $K_{\rm S}$      & Telescope \\
\hline
2003-06-12 & 52\,802.86           & 9.81$\pm$0.20    & \dots            & 7.48$\pm$0.20    & VLT   \\
2008-04-10 & 54\,567.78           & 12.63$\pm$0.16   & 10.11$\pm$0.09   & 8.16$\pm$0.14    & VLT   \\
2010-05-15 & 55\,271.34           & 17.65$\pm$0.04   & 15.09$\pm$0.09	& 12.27$\pm$0.07   & VISTA \\
2010-06-26 & 55\,373.52           & \dots	     & \dots	        & 12.69$\pm$0.09   & IRIS  \\
2010-06-27 & 55\,374.51           & $>$17.37	     & \dots		& \dots 	   & IRIS  \\
2010-06-28 & 55\,375.51           & \dots	     & \dots	        & 12.73$\pm$0.09   & IRIS  \\
2010-07-01 & 55\,378.51           & $>$16.98	     & 15.86$\pm$0.22	& 12.66$\pm$0.09   & IRIS  \\
2010-08-14 & 55\,423.01           & \dots	     & \dots		& 12.57$\pm$0.04   & VISTA \\
2011-05-12 & 55\,694.34           & \dots	     & \dots		& 12.38$\pm$0.03   & VISTA \\
2011-05-13 & 55\,695.40           & \dots	     & \dots		& 12.41$\pm$0.03   & VISTA \\
2011-05-15 & 55\,697.28           & \dots	     & \dots		& 12.41$\pm$0.04   & VISTA \\
2011-05-16 & 55\,698.34		  & \dots            & \dots            & 12.45$\pm$0.04   & VISTA \\
2011-05-29 & 55\,711.30		  & \dots	     & \dots		& 12.65$\pm$0.04   & VISTA \\
2011-06-08 & 55\,721.26		  & \dots	     & \dots		& 12.68$\pm$0.04   & VISTA \\
2011-08-17 & 55\,791.04		  & \dots	     & \dots		& 12.40$\pm$0.04   & VISTA \\
2011-08-22 & 55\,796.06		  & \dots	     & \dots		& 12.40$\pm$0.04   & VISTA \\
2011-08-31 & 55\,805.02		  & \dots	     & \dots		& 12.43$\pm$0.03   & VISTA \\
2011-09-01 & 55\,806.03		  & \dots	     & \dots		& 12.45$\pm$0.04   & VISTA \\
2011-09-19 & 55\,824.06		  & \dots	     & \dots		& 12.58$\pm$0.04   & VISTA \\
2011-10-02 & 55\,837.00		  & \dots	     & \dots		& 12.58$\pm$0.03   & VISTA \\
2012-06-04 & 56\,083.04		  & \dots	     & \dots		& 11.86$\pm$0.04   & VISTA \\
2012-06-04 & 56\,083.16		  & \dots	     & \dots		& 11.86$\pm$0.04   & VISTA \\
2012-06-20 & 56\,098.98		  & \dots	     & \dots		& 11.75$\pm$0.04   & VISTA \\
2012-06-20 & 56\,099.01		  & \dots	     & \dots		& 11.75$\pm$0.04   & VISTA \\
2012-06-20 & 56\,099.06		  & \dots	     & \dots		& 11.76$\pm$0.04   & VISTA \\
2012-06-26 & 56\,105.02		  & \dots	     & \dots		& 11.69$\pm$0.05   & VISTA \\
2012-06-26 & 56\,105.09		  & \dots	     & \dots		& 11.68$\pm$0.04   & VISTA \\
2012-06-29 & 56\,108.18		  & \dots	     & \dots		& 11.74$\pm$0.04   & VISTA \\
2013-05-26 & 56\,439.37		  & \dots	     & \dots		& 10.61$\pm$0.04   & VISTA \\
2013-06-26 & 56\,470.26		  & \dots	     & \dots		& 10.65$\pm$0.04   & VISTA \\
2013-06-27 & 56\,471.16		  & \dots	     & \dots		& 10.67$\pm$0.04   & VISTA \\
2013-06-29 & 56\,473.17		  & \dots	     & \dots		& 10.67$\pm$0.03   & VISTA \\
2013-06-30 & 56\,474.13		  & \dots	     & \dots		& 10.73$\pm$0.04   & VISTA \\
2013-07-01 & 56\,475.17		  & \dots	     & \dots		& 10.72$\pm$0.04   & VISTA \\
2013-07-13 & 56\,487.16		  & \dots	     & \dots		& 10.56$\pm$0.04   & VISTA \\
2013-07-17 & 56\,490.97		  & \dots	     & \dots		& 10.51$\pm$0.04   & VISTA \\
2013-07-20 & 56\,494.01		  & \dots	     & \dots		& 10.49$\pm$0.03   & VISTA \\
2013-07-20 & 56\,494.07		  & \dots	     & \dots		& 10.48$\pm$0.04   & VISTA \\
2013-07-20 & 56\,494.12		  & \dots	     & \dots		& 10.47$\pm$0.03   & VISTA \\
2013-07-20 & 56\,494.16		  & \dots	     & \dots		& 10.48$\pm$0.03   & VISTA \\
2013-07-20 & 56\,494.21		  & \dots	     & \dots		& 10.48$\pm$0.04   & VISTA \\
2013-07-21 & 56\,495.01		  & \dots	     & \dots		& 10.44$\pm$0.03   & VISTA \\
2013-07-21 & 56\,495.08		  & \dots	     & \dots		& 10.47$\pm$0.03   & VISTA \\
2013-07-21 & 56\,495.14		  & \dots	     & \dots		& 10.46$\pm$0.04   & VISTA \\
2013-07-21 & 56\,495.20		  & \dots	     & \dots		& 10.45$\pm$0.04   & VISTA \\
2013-07-24 & 56\,498.14		  & \dots	     & \dots		& 10.44$\pm$0.04   & VISTA \\
2013-07-27 & 56\,501.09		  & \dots	     & \dots		& 10.46$\pm$0.03   & VISTA \\
2013-07-30 & 56\,504.00		  & \dots	     & \dots		& 10.51$\pm$0.04   & VISTA \\
2016-06-07 & 57\,546.56		  & 15.07$\pm$0.03   & 12.67$\pm$0.02   & 10.22$\pm$0.03   & SMARTS 1.3\,m \\
2016-08-10 & 57\,610.52           & 14.44$\pm$0.03   & 12.17$\pm$0.02   & 9.75$\pm$0.04    & SMARTS 1.3\,m \\
2016-08-22 & 57\,622.51           & \dots	     & \dots            & 9.90$\pm$0.02    & IRIS \\
2016-08-25 & 57\,625.51           & 14.52$\pm$0.02   & 12.26$\pm$0.02   & 9.92$\pm$0.02    & IRIS \\

\hline
\end{tabular}
\end{table*}

\end{appendix}

\end{document}